\documentclass[twocolumn,showpacs,preprintnumbers,amsmath,amssymb,prl]{revtex4}
\usepackage{graphicx}

\begin{document}
\title{Fracture surfaces of heterogeneous materials: a 2D solvable model}
\author{E. Katzav, M. Adda--Bedia and B. Derrida}
\affiliation{Laboratoire de Physique Statistique de l'Ecole Normale Sup\'erieure, CNRS UMR 8550,\\
24 rue Lhomond, 75231 Paris Cedex 05, France.}
\date{\today}

\begin{abstract}

Using an elastostatic description of crack growth based on the Griffith criterion and the principle of local symmetry, we present a stochastic model describing the propagation of a crack tip in a 2D heterogeneous brittle material. The model ensures the stability of straight cracks and allows for the study of the roughening of fracture surfaces. When neglecting the effect of the non singular stress, the problem becomes exactly solvable and yields analytic predictions for the power spectrum of the paths. This result suggests an alternative to the conventional power law analysis often used in the analysis of experimental data.

\end{abstract}


\maketitle

The field of fractography is concerned with the characterization and study of surfaces produced when solids break \cite{Hull}. In contrast to other pattern formation mechanisms governed by a phase transition or material deposition, in fracture phenomena the broken surfaces are created by irreversible material separation induced by the propagation of a crack front. A common characteristic of many materials is that these surfaces are not smooth. Concerning this feature, the pioneering work of Mandelbrot {\em et al.} \cite{Mandelbrot84} and further developments \cite{dev, Plouraboue96, Balankin05, Paper} have shown that crack surfaces can also exhibit fractal characteristics. For a one dimensional curve, say $h(x)$, embedded in 2D geometry, this self-affine property is characterized by an exponent $\zeta$, called the roughness exponent. There are many ways to quantify roughness \cite{Schmittbuhl95a}, one of them being based on the power-spectrum $P(q)= \langle h_q h_{-q}\rangle$. For a self-affine path, it is expected that $P(q) \sim q^{-1-2\zeta}$ for intermediate, yet small, $q$'s. Measurements of the roughness exponent of crack paths in quasi-two-dimensional materials have been performed on Berea sandstone ($\zeta \simeq 0.8$) \cite{Plouraboue96}, concrete ($\zeta\simeq 0.75$) \cite{Balankin05} and paper ($\zeta\simeq 0.6$--$0.7$) \cite{Paper}. These variations question the concept of universality of fracture roughness exponent in 2D geometries.

A theoretical description of the fracture surface must follow from the study of crack tip propagation. A key ingredient for that comes from the divergence of the elastostatic stress field in the vicinity of the crack tip
\begin{equation}
\sigma_{ij}(r,\theta)=\sum_{\ell=1,2}\frac{K_\ell}{\sqrt{2\pi r}}\,\Sigma^{(\ell)}_{ij}(\theta)+T\delta_{ix}\delta_{jx}+O\left(\sqrt{r}\right)\;,
\label{eq:SIF}
\end{equation}
where $(r,\theta)$ are polar coordinates with $r=0$ located at the crack tip, the $x$-direction is parallel to the extension of the crack path at $r=0$, and $\Sigma^{(\ell)}_{ij}$ are known functions describing the angular variations of the stress field components \cite{Broberg}. In this expansion, $K_\ell$ ($\ell=1,2$) and $T$ are the static stress intensity factors (SIFs) and the nonsingular $T$-stress, respectively. In the framework of linear elastic fracture mechanics, the crack propagation is governed by the Griffith energy criterion and the Principle of Local Symmetry (PLS) \cite{Broberg}. In its simplest form, the Griffith criterion states that the crack must grow by satisfying $K_1 = K_{1c}$, where $K_{1c}$ is the material toughness. The PLS, formulated as $K_2=0$, imposes a pure opening mode for the stress field in the vicinity of the crack tip. Within these two criteria, the crack path is mainly selected by the PLS, while Griffith's criterion controls the intensity of the loading necessary for the crack to grow. In addition, the nonsingular $T$-stress in Eq.~(\ref{eq:SIF}) rules the stability of crack propagation. Stability analysis around a straight crack path under pure tensile loading has shown that if $T>0$ straight crack propagation is unstable, while stable when $T < 0$ \cite{Cotterell80}.

\begin{figure}[ht]
\centerline{{\includegraphics[width=7cm]{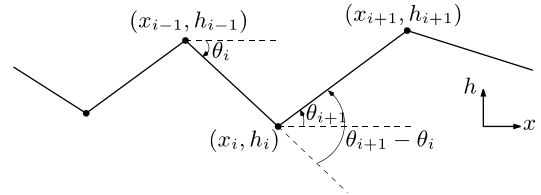}}}
\caption{Modeling crack propagation in a heterogeneous medium as a succession of smooth parts and kinks due to local defects.}
\label{fig:model}
\end{figure}

In this work, we model crack propagation in 2D brittle heterogeneous media by using the above elastostatic approach formulated in terms of the SIFs and the $T$-stress. We suppose that the crack propagates smoothly by satisfying the Griffith criterion and the PLS until it meets a heterogeneity which induces a change in the fracture toughness and/or a local shear perturbation. In response, the crack changes locally its direction of propagation by forming a kink. If the distance between kinking events is small one can show that at leading order the crack extensions are straight \cite{Leblond92}. As shown in Fig.~\ref{fig:model}, the local kinking angle at the position $i$ is given by $\delta\theta_i=\theta_{i+1}-\theta_{i}$, which in the continuum limit is $\delta\theta(x)=h''(x)\delta x$. Now, in order to write an equation for the crack path $h(x)$ one should relate $\delta \theta (x)$ to the material heterogeneities and to the elastic quantities.

Consider a crack in a material subjected to a loading such that in the absence of defects, a straight line $h(x)=0$ is the solution for the crack path. For this configuration, the variation of the kinking angle $\delta\theta$ is related to the variation of the mode~II stress intensity factor $\delta K_2$ induced by the defects \cite{Leblond92,Sumi92}. To linear order, this relation is given by $\delta \theta = -2(\delta K_2)/K_{1}$ \cite{Leblond92,Sumi92}. Next, one can write $\delta K_2 (\{h\},x)$ as a functional of $h(t)$, the shape of the crack path for $t\leq x$ \cite{MovchanGao}. Putting all together, one obtains the following dimensionless equation for the crack path
\begin{equation}
h''(x)=\eta_2'(x) + \eta_1'(x)\left[ h'(x)+\alpha h(x)-\beta \int_{-\infty}^x \frac{h'(t)dt}{\sqrt{x-t}}  \right] ,
\label{eq:start}
\end{equation}
A systematic derivation of this equation will be given in \cite{detailed}. In Eq.~(\ref{eq:start}), the primes denote derivatives, $\alpha$ is a positive quantity fixed by the geometry of the sample in which the crack propagates \cite{MovchanGao} and $\beta$ is proportional to the $T$-stress. In addition, Eq.~(\ref{eq:start}) contains two space dependent functions which represent the local fluctuations in the toughness, modeled by $\eta_1(x)$, and local shear fluctuations, modeled by $\eta_2(x)$. Note that when the material is homogeneous, $\eta'_\ell(x)=0$ ($\ell=1,2$), and the solution of Eq.~(\ref{eq:start}) is given by a straight path, recovering the zeroth order solution. Therefore Eq.~(\ref{eq:start}) describes the deviation from a straight crack propagation in the presence of inhomogeneities.

\begin{figure}[ht]
\centerline{\includegraphics[width=7cm]{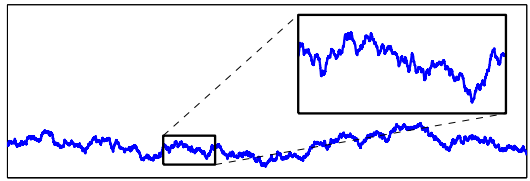}}
\caption{An example of a crack path produced using Eq.~(\ref{eq:start-dis}), with $\Delta=10^{-5}$, $\alpha=20$, $D_1=0.04$ and $D_2=6.25$. The inset shows a zoom of a smaller part of the path.}
\label{fig:zoom}
\end{figure}

The numerical resolution of Eq.~(\ref{eq:start}) reveals that when $\beta > 0$ the crack paths are unstable, while they are stable for $\beta \le 0$ \cite{detailed}. This is consistent with the $T$-criterion \cite{Cotterell80}, however we found that for $\beta \le 0$ the crack propagation is more stable in our formulation, since the crack path decays to a straight line rather than exhibits a square-root behavior ($h(x)\sim \sqrt{x}$) \cite{Cotterell80}. In the following, we focus on the case $\beta=0$ for two reasons. First, power counting shows that as long as the path is stable the $\beta$-term is subdominant at large scales. This is verified by numerical simulations and turns out to be valid at all scales \cite{detailed}. Second, the model is exactly solvable for this case, which allows for conclusive predictions. The discrete version of Eq.~(\ref{eq:start}) with $\beta=0$ yields
\begin{eqnarray}
h_{i+1}&=&2h_i-h_{i-1} +(\eta_{2,i}-\eta_{2,i-1})\Delta \nonumber\\
&&+(\eta_{1,i}-\eta_{1,i-1})\left( h_i-h_{i-1}+\alpha h_i \Delta\right)\;,
\label{eq:start-dis}
\end{eqnarray}
where $1<i\leq N$. The quantities $\eta_{\ell,i}$ are modeled as independent random numbers, characterized by their variances $\left \langle \eta_{\ell,i} \eta_{m,j} \right \rangle = D_\ell \delta_{\ell m} \delta_{ij}$. The initial conditions are chosen such that $h(x)=0$ for $x \le 0$, implying $h_0=h_1=0$. In addition, the $x$-axis is scaled such that the observation window is always $0<x<1$, thus for a $1D$ lattice with $N$ sites, one has $\Delta = 1/N$. Fig.~\ref{fig:zoom} shows an example of a path grown using this procedure and shows that the model produces rough crack lines with possible self-similar properties.

\begin{figure}[ht]
\centerline{\includegraphics[width=4.5cm]{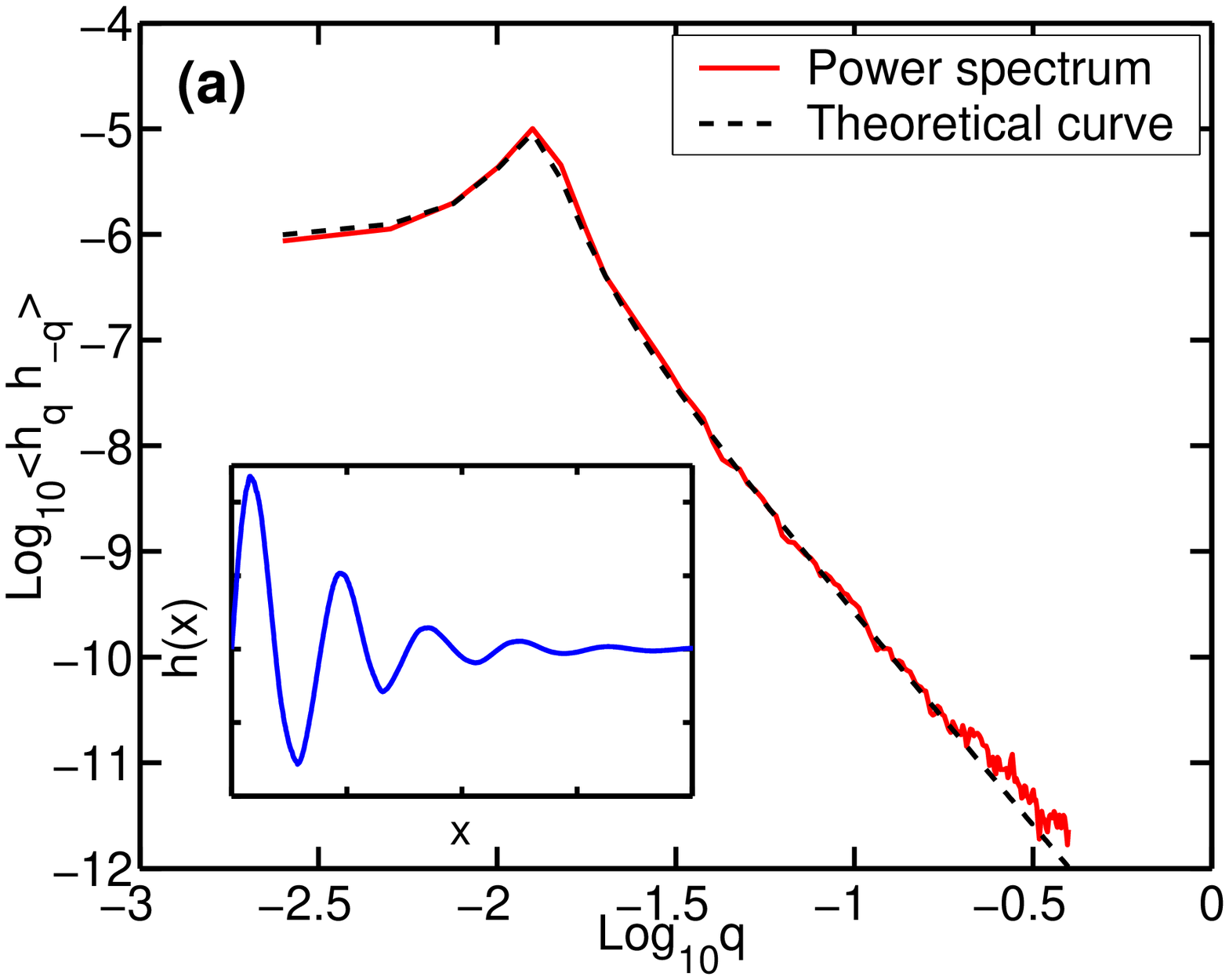}\includegraphics[width=4.3cm]{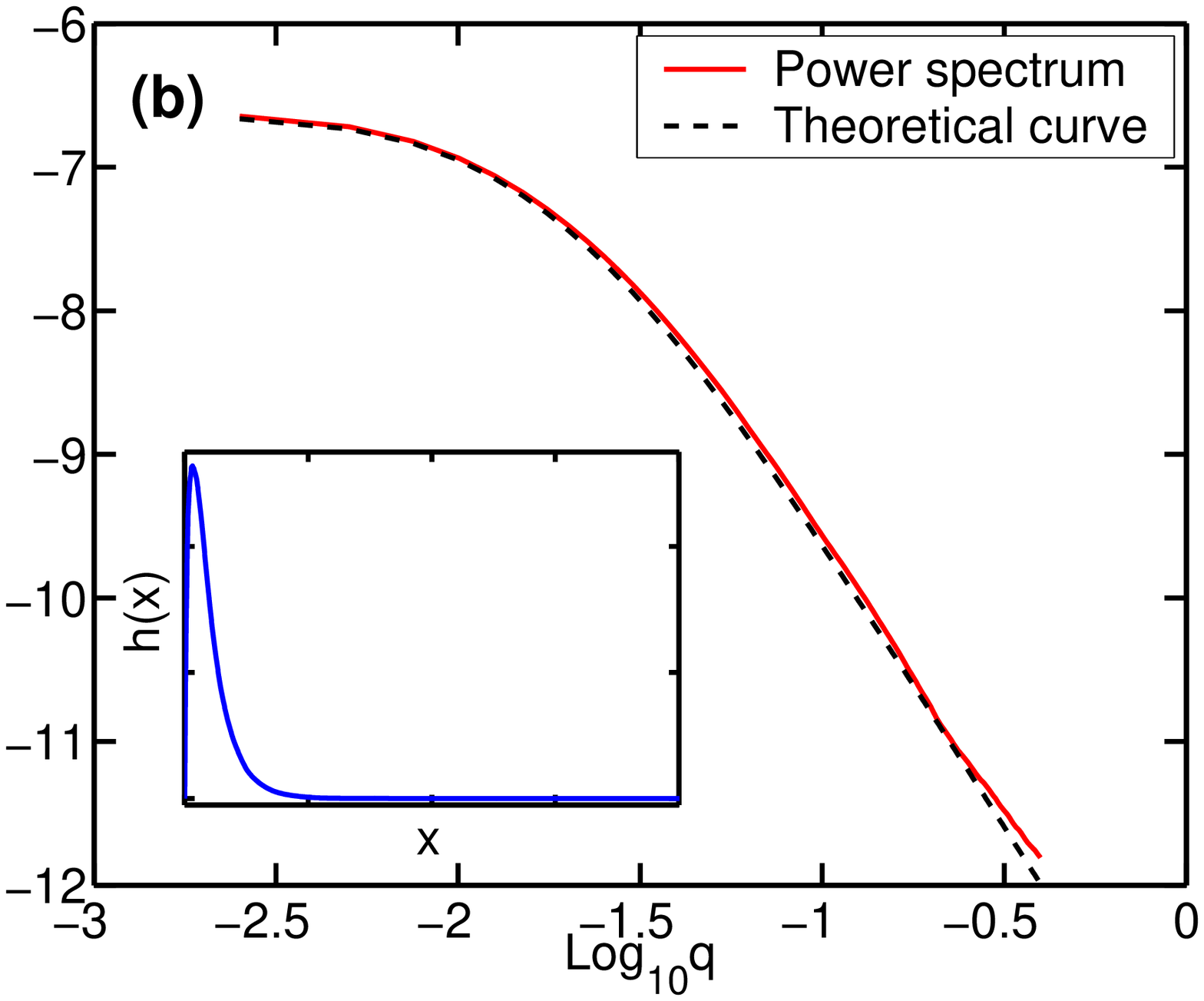}}
\caption{Power spectra of crack paths with a single shear perturbation averaged over $10$ realizations, using $\theta_0=1$ and $\Delta=4\times10^{-4}$. (a) $\alpha=100$, $D_1=4\times10^{-3}$ ($C<4\alpha$). (b) $\alpha=20$, $D_1=4\times10^{-2}$ ($C>4\alpha$). The solid lines correspond to the theoretical curves. The insets show examples of such paths.}
\label{fig:eta1}
\end{figure}

Let us first study crack paths when there is a single shear perturbation given by $\eta_{2,i}=\theta_0 \delta_{i2}$. This situation is equivalent to redefining the boundary conditions as $h(0)=0$ and $h'(0)=\theta_0$ \cite{Cotterell80}. Later, we will generalize our results to any uncorrelated shear perturbations $\eta_{2,i}$ and find that this source of randomness is essential to reproduce the experimentally observed rough crack paths. For $\eta_{2,i}=\theta_0 \delta_{i2}$, we are going to show that crack paths tend to decay into straight ones (see insets in Fig.~\ref{fig:eta1}). This decay can be also accompanied by oscillations. Interestingly, when considering an ensemble of paths grown using the same parameters $\theta_0$, $\alpha$ and $D_1/\Delta$, but with different choices of the noise $\eta_{1}$ one finds that for small $\Delta$, the solution becomes independent of the realization of the noise. To understand this, let us consider the logarithmic derivative of $h$, namely $\psi_i=\frac{h_i-h_{i-1}}{h_i\Delta}$ \cite{Derrida84}. In terms of $\psi_i$, Eq.~(\ref{eq:start-dis}) with $\eta_{2,i}=0$ can be rewritten as
\begin{equation}
\psi_{i+1} = \alpha \left(\eta_{1,i} -\eta_{1,i-1}\right)+\frac{\psi_i + \left(\eta_{1,i} -\eta_{1,i-1}\right)\psi _i}{1 + \psi_i\Delta} \;.
\label{eq:start-psi}
\end{equation}
From the observation that the solutions depend on $D_1/\Delta$ only, $\eta_{1}$ should be of order $O (\sqrt{\Delta})$. We will be interested in deriving a Fokker-Planck equation for the $\psi$'s. As usual in stochastic equations, the analysis of Eq.~(\ref{eq:start-psi}) for small $\Delta$ requires keeping terms up to order $\Delta$, since the leading order (order $\sqrt\Delta$) has zero average. This gives
\begin{equation}
\psi_{i+1}= \psi_i + \left(\eta_{1,i} - \eta_{1,i-1}\right) \left(\psi_i + \alpha \right) - \psi_i^2 \Delta \;.
\label{eq:start-psi1}
\end{equation}
Next, using Eq.~(\ref{eq:start-psi1}) one writes the integral equation for the probability distribution of the $\psi$'s, namely $P(\psi,\eta_1)$, which is readily given by
\begin{eqnarray}
&&P_{i+1}\left(\psi_{i+1},\eta_i\right) = \int{d\eta_{i-1}\rho\left({\eta_{i-1}}\right)\int{d\psi_i P_i\left(\psi_i,\eta_{i-1}\right)}} \nonumber \\
&&\delta \left(\psi_{i+1}-\psi_i-\left({\eta_i-\eta_{i-1}}\right)\left({\psi_i+\alpha}\right)+\psi_i^2\Delta\right)\; ,
\label{eq:FP1}
\end{eqnarray}
where $\rho(\eta_i)$ is the distribution of the random toughness at the location $i$. For brevity, the subscript $1$ in $\eta_{1}$ has been suppressed. In order to integrate over $\psi_i$ and to eliminate the $\delta$ function from the integral in Eq.~(\ref{eq:FP1}), one needs to express $\psi_i$ as a function of $\psi_{i+1}$. To achieve that we iterate Eq.~(\ref{eq:start-psi1}) to first order in $\Delta$ and then get
\begin{widetext}
\begin{equation}
P_{i+1}\left({\psi_{i+1},\eta_i}\right)=\int{d\eta_{i-1}\rho\left({\eta_{i-1}}\right)\frac{{P_i\left({\psi_{i+1}-\left({\eta_i-\eta_{i-1}}\right)\left({\psi_{i+1}+\alpha}\right)+\left({\eta_i-\eta_{i-1}}\right)^2\left({\psi_{i+1}+\alpha}\right)+\psi_{i+1}^2\Delta,\eta_{i-1}}\right)}}{{1+\left({\eta_i-\eta_{i-1}}\right)-2\psi_{i+1}\Delta}}}\;.
\label{eq:FP2}
\end{equation}
\end{widetext}
The denominator in Eq.~(\ref{eq:FP2}) results from the Jacobian of the $\delta$-function. Expanding the integrand to order $\Delta$ gives
\begin{widetext}
\begin{eqnarray}
 P_{i+1}\left({\psi,\eta_i}\right) &=& \int{d\eta_{i-1}\rho\left({\eta_{i-1}}\right)\left\{{P_i\left({\psi,\eta_{i-1}}\right)-\left({\eta_i-\eta_{i-1}}\right)\frac{d}{{d\psi}}\left[{\left({\psi+\alpha}\right)P_i\left({\psi,\eta_{i-1}}\right)}\right]}\right.}\nonumber\\
 && \left.{+\left({\eta_i-\eta_{i-1}}\right)^2\frac{{d^2}}{{d\psi^2}}\left[{{\textstyle{1\over2}}\left({\psi+\alpha}\right)^2P_i\left({\psi,\eta_{i-1}}\right)}\right]+\Delta\frac{d}{{d\psi}}\left[{\psi^2P_i\left({\psi,\eta_{i-1}}\right)}\right]}\right\}
 \label{eq:FP3} \ .
\end{eqnarray}
\end{widetext}

We see that, at order $\Delta$, the probability distribution function $P_{i+1}$ is quadratic in $\eta_i$. Thus one can write
\begin{equation}
P_{i}(\psi,\eta_{i-1})=P_{i}^{(0)}(\psi)+\eta_{i-1} P_{i}^{(1)}(\psi)+\eta_{i-1}^2P_{i}^{(2)}(\psi)\;.
\end{equation}
Plugging this expansion into Eq.~(\ref{eq:FP3}) and performing the integral over $\eta_{i-1}$ gives $P_{i+1}^{(1)}\left(\psi\right)=-\frac{d}{{d\psi}}\left[{\left({\psi+\alpha}\right)P_i^{(0)}\left(\psi\right)}\right]$ and $P_{i+1}^{(2)}\left(\psi\right)=\frac{{d^2}}{{d\psi^2}}\left[{\frac{1}{2}\left({\psi+\alpha}\right)^2P_i^{(0)}\left(\psi\right)}\right]$, and finally we get from Eq.~(\ref{eq:FP3}) the $x$-dependent Fokker-Planck equation for $P^{(0)}$
\begin{equation}
\frac{\partial P^{(0)}}{\partial x} = \frac{d}{d\psi} \left[ \left(\psi^2 + C\left( {\psi  + \alpha } \right) \right)P^{(0)}\right]\;,
\label{eq:FP8}
\end{equation}
where $C\equiv\frac{D_1}{\Delta}$, with $D_1=\left\langle\eta_1^2\right\rangle$. Note that Eq.~(\ref{eq:FP8}) does not have second derivatives of $P^{(0)}$ with respect to $\psi$. Thus, strictly speaking, it looks like a Liouville equation. This means that the probability distribution function of $\psi$ becomes that of a deterministic evolution given by
\begin{eqnarray}
&&P_{i+1}(\psi_{i+1}) = \int d\psi_iP_i(\psi_i)\nonumber\\
&&\delta\left(\psi_{i+1}-\psi_i+\left[\psi_i^2 + C (\psi_i  + \alpha)\right]\Delta\right)\;,
\label{eq:liouville}
\end{eqnarray}
from which we can read the deterministic equation
\begin{equation}
\frac{\partial \psi}{\partial x} = -\psi^2 - C \left( \psi + \alpha \right)\;.
\label{eq:psi-c}
\end{equation}
Since in the continuum limit $\psi(x)=h'(x)/h(x)$, Eq.~(\ref{eq:psi-c}) yields
\begin{equation}
h''(x) = -C\left[ h'(x) + \alpha h(x) \right] \;.
\label{eq:h-c}
\end{equation}
Thus, one finds that the effective equation (\ref{eq:h-c}) is obtained from the noisy one (Eq.~(\ref{eq:start}) with $\beta=0$) by simply replacing the noise term $\eta_1'(x)$ by a negative constant, $-C$, proportional to its variance, even though $\eta_1'(x)$ is equally positive and negative. This nontrivial result shows that, in the small $\Delta$ limit, the local toughness fluctuations $\eta_{1,i}$ do not influence the shape of the crack path, apart from setting a relevant scale for the energy that has to be invested in making the crack propagate. At this point, one can easily solve Eq.~(\ref{eq:h-c}) with the initial conditions $h(0)=0$ and $h'(0)=\theta_0$:
\begin{equation}
h(x) = \theta_0e^{-\frac{1}{2} Cx} \frac{\sinh \left(\frac{1} {2}\sqrt {C^2 - 4\alpha C} x \right)}{\frac{1}{2} \sqrt {C^2 - 4\alpha C}} \;.
\label{eq:h-csol}
\end{equation}
The solution for the crack path (\ref{eq:h-csol}) exhibits either an exponential decay or damped oscillations depending of the sign of $(C-4\alpha)$. For the present case, the power spectrum of the crack path can be obtained analytically. It is given by $\left\langle h_q h_{-q}\right\rangle = \theta_0^2\left[\left(q^2 - C\alpha \right)^2 + C^2 q^2\right]^{-1}$, where $h_q$ is the Fourier component of $h(x)$. In Fig.~\ref{fig:eta1}, we compare this expression for $\left\langle h_q h_{-q}\right\rangle $ with the corresponding averaged power spectrum over ten realizations of the noise $\eta_1$. As can be seen, both the cases $C<4\alpha$ (damped oscillations) and $C>4\alpha$ (exponential decay) show that the theoretical results fit the numerical data.

\begin{figure}[ht]
\centerline{\includegraphics[width=4.5cm]{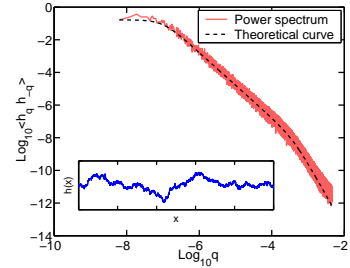}}
\caption{Power spectrum of crack paths with many shear perturbations averaged over $10$ realizations, using $\Delta=10^{-4}$, $\alpha=1$, $D_1=0.04$ and $D_2=4.2$. The solid line corresponds to the theoretical prediction. The inset shows an example of such a path.}
\label{fig:eta2}
\end{figure}

Our next step is to consider crack propagation when there are many shear perturbations. This amounts to retaining the additive noise $\eta_2$ in Eq.~(\ref{eq:start-dis}). Unlike the fluctuations in the local toughness $\eta_1$, the shear perturbations cannot be modeled by a deterministic term. Furthermore, its randomness seems crucial for creating the rough patterns similar to those observed in experiments \cite{Plouraboue96,Balankin05,Paper} (see for example Fig.~\ref{fig:zoom}). Interestingly, by varying the amplitude of the noise $D_2$, one can can produce rather different patterns. Based on the previous results we generalize the expression for the power-spectrum to the current case. Note that in Eq.~(\ref{eq:start-dis}), $\eta_2$ is just a non homogeneous term. Thus, once we have a solution for the homogeneous equation, we can construct a special solution for the non homogeneous one. Using the fact that $\eta_1$ and $\eta_2$ are independent random variables, we conclude that averaging over realizations of the local toughness fluctuations still amounts to replacing $\eta_1'(x)$ by $-C$. Doing so, we write the power spectrum for a general noise $\eta_2(x)$ in the form
\begin{equation}
\left \langle h_q h_{-q} \right \rangle = D_2 \frac{\left(1+C^{-1}\right) q^2 + (2C+\alpha+2)} {\left(q^2 - C\alpha \right)^2 + C^2 q^2}\;,
\label{eq:h2-cPS}
\end{equation}
which is different from the power-spectrum in the case of a single
shear perturbation, as there is a $q^2$ term in the numerator. The
coefficients in Eq.~(\ref{eq:h2-cPS}) are determined from the
Fourier transform of a stationary signal after cutting out the
transient regime, that is $h_q=\int_{x_0}^1 h(x) e^{iqx}dx$. This
leads to a system with effective random initial conditions at
$x=x_0$. Since $x_0$ is chosen in the steady regime the statistics
of $h(x_0)$ and $h'(x_0)$ are known, and an analytical expression of
the power spectrum can be obtained. Fig.~\ref{fig:eta2} shows that
despite Eq.~(\ref{eq:h2-cPS}) agrees very well with the simulations
data for the power spectrum, one might be tempted to fit it with a
power law ansatz. However, apart from a tail $q^{-2}$ at large
$q$'s, which yields a roughness exponent $\zeta=0.5$ at small length
scales, Eq.~(\ref{eq:h2-cPS}) tells us that there is no self-affine
behavior of fracture surfaces at intermediate length scales.

Here we have introduced a model that describes the motion of a crack tip in a 2D heterogeneous brittle material and allows for the analysis of the form of the surface created. The approach is based on a pure elastic description of the crack growth and does not necessitate additional ingredients such as propagation through voids nucleation \cite{Eytan05}. The results for the stability of a straight crack propagation are in agreement with the well known $T$-criterion \cite{Cotterell80}. Exact results are provided for the case $\beta=0$, that allow to predict analytically the power spectrum of the crack surface. These results suggest that the roughness of the surface is not self-affine, but rather a fuzzy shape convoluted with an oscillating path.

From an experimental point of view, it could be interesting to use expressions like Eq.~(\ref{eq:h2-cPS}) instead of simple power-laws to fit the data of rough crack lines in 2D geometries. However, such analysis require averaging over many repetitions under the same experimental situations, which is difficult to perform due to the irreversibility of fracture events. Finally, our conclusions are not relevant to crack surfaces only, but concern analysis of rough shapes where an oscillating background is likely to exist and might bias the measurement of the surfaces characteristics \cite{Schmittbuhl95a,detailed}.

This work was supported by EEC PatForm Marie Curie action (E.K.).

\end{document}